\newcommand{\be}{\begin{equation}}
\newcommand{\ee}{\end{equation}}
\newcommand{\bea}{\begin{eqnarray}}
\newcommand{\eea}{\end{eqnarray}}
\newcommand{\nn}{\nonumber}
\def\beq{\begin{equation}}
\def\eeq{\end{equation}}
\def\bea{\begin{eqnarray}}
\def\eea{\end{eqnarray}}
\def\nn{\nonumber}

\def\la{\lambda} \def\lap{\lambda^{\prime}}  \def\de{\delta}  \def\dag{\dagger}
    
\def\bnabla{\boldmath{\nabla}}
\def\bm{\mathbf}
\def\rm{\textrm}

\documentclass{PoS}

\title{EFT determination of the heavy-hybrid spin potential}

\ShortTitle{EFT determination of the heavy-hybrid spin potential}

\author{\speaker{Wai Kin Lai}\thanks{We thank the organizers of Confinement 2018 for the invitation to give a talk in the parallel session.}
       \\
        Physik-Department, Technische Universit\"at M\"unchen, \\
        James-Franck-Stra\ss e 1, 85748 Garching, Germany\\
        E-mail: \email{wk.lai@tum.de}}

\abstract{We study the spin splitting in the heavy quarkonium hybrid spectrum within the framework of
an nonrelativistic effective field theory. We derive for the first time the spin-dependent part of 
the heavy-quark-antiquark potential for heavy quarkonium hybrids to order $1/m^2$ in the heavy-quark-mass expansion. We find that 
several operators that are not found in standard quarkonia appear, most remarkably
an operator suppressed by only one power
of the heavy-quark mass. 
By matching the weakly-coupled pNRQCD to the effective field theory in the regime
of short heavy-quark-antiquark distances,
we work out the matching coefficients of the spin-dependent operators, which are factorized into 
a perturbative and a nonperturbative part. The nonperturbative part can be expressed in
terms of purely gluonic correlators. We fit the nonperturbative parts of the matching
coefficients to lattice data of the charmonium hybrid spectrum and obtain
results that respect the power counting. 
Using the obtained nonperturbative pieces, 
we compute the bottomonium hybrid spectrum with the spin-dependent potential, for which results from the lattice are still sparse.}

\FullConference{XIII Quark Confinement and the Hadron Spectrum - Confinement2018\\
		31 July - 6 August 2018\\
		Maynooth University, Ireland}

\begin{document}
\section{Introduction}
Since the discovery of the $X(3872)$ 
by the Belle Collaboration in 2003~\cite{Choi:2003ue},
more than two dozens of nontraditional charmonium- and bottomonium-like states, 
the so-called XYZ mesons, have been observed at B-factories
(BaBar, Belle and CLEO), $\tau$-charm facilities (CLEO-c, BESIII) and also proton-(anti)proton colliders 
(CDF, D0, LHCb, ATLAS, CMS). There is strong evidence that some of these states are non-conventional hadrons, in the sense 
that they do not fit into the quark-model picture. Various theoretical interpretations of these exotic hadrons have been proposed and studied 
(see, e.g., the reviews~\cite{Brambilla:2004wf,Brambilla:2010cs, Brambilla:2014jmp,Olsen:2014qna}
for more details on the experimental and theoretical status of the subject).
One attractive interpretation of the XYZ mesons is the heavy quarkonium hybrid, which contains a heavy quark and a heavy antiquark, together with a gluonic
excitation. While other interpretations of the XYZ mesons are multiquark states, which under some circumstances have an electromagnetic analogue in molecular states,
quarkonium hybrids, having a gluonic excitation, shows a unique feature of QCD.

There are two approaches fully rooted in QCD to the computation of the spectra of quarkonium hybrids:
lattice simulations and effective field theories (EFTs). 
Studies of heavy hybrids on the lattice have traditionally focused on the charmonium sector. 
Calculations in the bottomonium sector are more challenging since smaller lattice spacings are required.
A pioneering quenched calculation of the excited charmonium spectrum was presented in Ref.~\cite{Dudek:2007wv}.
This study was extended by the RQCD collaboration~\cite{Bali:2011dc, Bali:2011rd} and by the Hadron Spectrum Collaboration~\cite{Liu:2012ze, Cheung:2016bym}. 
The common result in all these studies has been the identification of the lowest hybrid charmonium spin-multiplet at about $4.3\,\textrm{GeV}$ 
containing a state with exotic quantum numbers $J^{PC}=1^{-+}$.

The EFT approach exploits the hierarchy of scales in quarkonium hybrids: $m\gg mv\gg \Lambda_{\rm QCD}\gg mv^2$, where $m$ is the mass of the heavy quark and $v$ the relative velocity between the heavy quark and antiquark. 
Most remarkably is $\Lambda_{\rm QCD}\gg mv^2$, as can be deduced from the lattice data, which show that an energy gap exists between the gluonic excitations and the excitations of the heavy-quark-antiquark pair~\cite{Juge:1997nc,Bali:2000vr,Juge:2002br,Bali:2003jq}. Note that the nonperturbative gluon dynamics occurs at the scale $\Lambda_{\rm QCD}$.
As a result, the heavy-quark-antiquark pair binds in the background potential created by the gluonic excitations, thus justifying 
the Born--Oppenheimer approximation~\cite{Griffiths:1983ah,Juge:1997nc,Braaten:2014qka,Braaten:2014ita}. Computation of the heavy hybrid spectrum
using a rigorous and systematic EFT approach has been performed~\cite{Berwein:2015vca,Oncala:2017hop,Brambilla:2017uyf}. In the EFT approach,
the interquark potential factorizes into perturbative and nonperturbative contributions.
The former can be calculated in perturbation theory, while the latter are parametrized according to the power counting, and can
in principle be calculated on the lattice.
 
In this paper, we report our study of the spin splitting in the quarkonium hybrid spectrum using the EFT approach~\cite{Brambilla:2018pyn}. In Section~\ref{Formalism}, we construct the EFT for the quarkonium hybrids at the scale $mv^2$ by integrating out the scale $\Lambda_{\rm QCD}$ in weakly-coupled pNRQCD in the regime of short interquark distances, obtaining the spin-dependent potential in the hybrid EFT to $\mathcal{O}(1/m^2)$. The spin-dependent potential
is shown to be parametrized by several nonperturbative parameters, which can be expressed in terms of purely gluonic correlators (details to be published in~\cite{Long_spin}). 
In Section~\ref{Results}, we fit the nonperturbative parameters in the EFT to lattice data of the charmonium hybrid spectrum. From the obtained nonperturbative parameters, we predict the 
the bottomonium hybrid spectrum. We conclude in Section~\ref{Conclusions}.
\section{Derivation of the spin-dependent potential}\label{Formalism}
\begin{figure}[ht]
\centerline{\includegraphics[width=.99\textwidth]{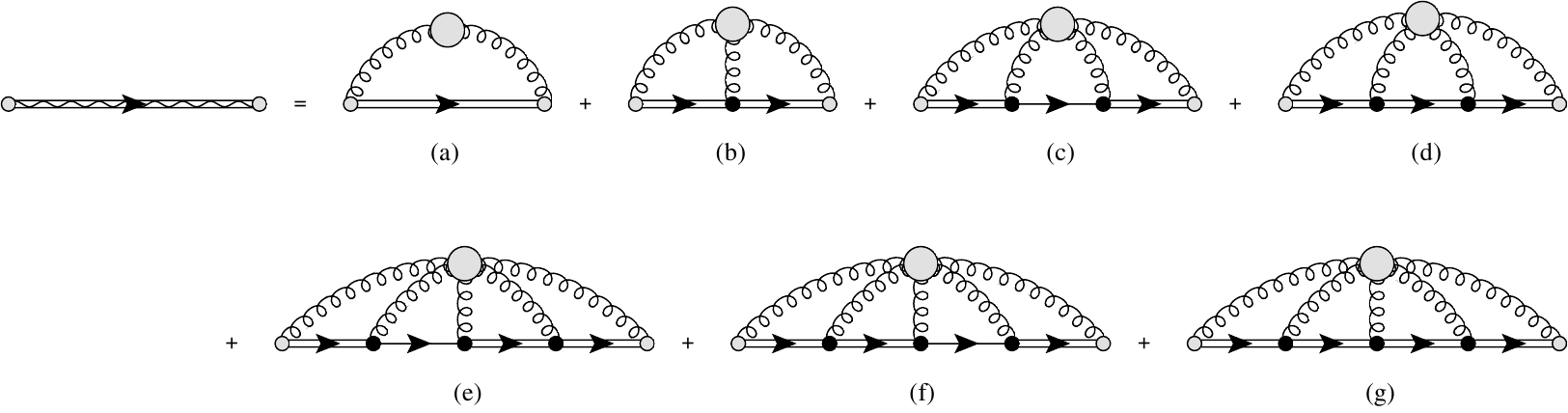}}
\caption{Matching of two-point function in the hybrid EFT on the left-hand side to weakly-coupled pNRQCD on the right-hand side. The single, double and curly lines represent the heavy-quark singlet, heavy-quark octet and gluon fields respectively. The black dots stand for vertices from weakly-coupled pNRQCD that involve a chromoelectric or chromomagnetic field.
The shaded blobs represent the nonperturbative gluon dynamics.}
\label{match}
\end{figure}

The first step in constructing the quarkonium hybrid EFT is to integrate out the scale $m$, which produces the well-known EFT called NRQCD \cite{Caswell:1985ui,Bodwin:1994jh,Manohar:1997qy}. The next step is to integrate out the scale $mv$. In the regime of short interquark distances, this step can be performed in perturbation theory and one arrives at the so-called weakly-coupled pNRQCD \cite{Pineda:1997bj,Brambilla:1999xf}. To arrive at the hybrid EFT at the scale $mv^2$, we perform a matching between weakly-coupled pNRQCD and the hybrid EFT in the short-distance regime 
$r\ll 1/\Lambda_{{\rm QCD}}$. First, we have to define the degrees of
freedom in the hybrid EFT.  
Following~\cite{Berwein:2015vca} we classify quarkonium hybrids according to their behavior at $r\rightarrow 0$
in the static limit.
In this limit, quarkonium hybrids reduce to gluelumps, which are bound states made of a color-octet heavy-quark-antiquark pair 
and some gluonic fields in a color-octet configuration 
localized at the center of mass of the heavy-quark-antiquark pair~\cite{Foster:1998wu,Brambilla:1999xf,Berwein:2015vca}.
A basis of gluelump states can be written as 
\bea
|\kappa,\,\lambda\rangle =P^i_{\kappa\lambda} \, O^{a\,\dagger}\left(\bm{r},\bm{R}\right) \, G_{\kappa}^{ia}(\bm{R})|0\rangle\,,
\label{eigen1}
\eea
where $O^a$ is the field operator for a color-octet heavy-quark-antiquark pair in pNRQCD.
$\bm{r}$ and $\bm{R}$ are the relative and center-of-mass coordinates of the heavy-quark-antiquark pair. 
$G^{ia}_\kappa$ is the operator that creates a gluonic excitation with quantum number $\kappa=K^{PC}$, with $\bm{K}$ the angular momentum of the gluonic degrees of freedom. 
$P^i_{\kappa\lambda}$ is the projector that projects the gluonic degrees of freedom to an eigenstate of $\bm{K}\cdot\hat{\bm{r}}$ with eigenvalue $\lambda$. 
The states $|\kappa,\,\lambda\rangle$ live in representations, characterized by
$\lambda$, of the cylindrical-symmetry group $D_{\infty h}$ (with $P$ replaced by $CP$), 
which is the same symmetry group of diatomic molecules.
The degrees of freedom
in the quarkonium hybrid EFT are the fields $\Psi_{\kappa\lambda}(t,\,\bm{r},\,\bm{R})$ associated to the states $|\kappa,\,\lambda\rangle$~\cite{Berwein:2015vca,Brambilla:2017uyf}. 
The Lagrangian of the quarkonium hybrid EFT resulting from the matching has the form
\bea
L_{\textrm{hybrid}} = \int d^3Rd^3r \, \sum_{\kappa} \sum_{\lambda\lambda^{\prime}}\Psi^{\dagger}_{\kappa\lambda}(\mathbf{r},\,\mathbf{R},\,t) \biggl\{i\partial_t - V_{\kappa\lambda\lambda^{\prime}}(r)+ P^{i\dag}_{\kappa\lambda}\frac{\bnabla^2_r}{m}P^i_{\kappa\lambda^{\prime}}\biggr\}\Psi_{\kappa\lambda^{\prime}}(\mathbf{r},\,\mathbf{R},\,t)+\dots\,,\,\,\,\,\,\,\,\,
\label{bolag2}
\eea
where the ellipsis stands for operators producing transitons to standard quarkonium states and transitions between hybrid states of different $\kappa$. The former are beyond the scope of this work and the latter are suppressed. 
The potential $V_{\kappa\la\lap}(r)$ can be expanded in $1/m$ as 
\bea
V_{\kappa\la\lap}(r)&=&V^{(0)}_{\kappa\la}(r)\de_{\la\lap}+\frac{V^{(1)}_{\kappa\la\lap}(r)}{m}+\frac{V^{(2)}_{\kappa\la\lap}(r)}{m^2}+\dots\,. 
\eea
We will write
\bea
V_{\kappa\la\lap}^{(1)}(r)&=&V_{\kappa\la\lap\,{\rm SD}}^{(1)}(r)+V_{{\kappa\la\lap}\,{\rm SI}}^{(1)}(r)\,,\\
V_{\kappa\la\lap}^{(2)}(r)&=&V_{\kappa\la\lap\,{\rm SD}}^{(2)}(r)+V_{{\kappa\la\lap}\,{\rm SI}}^{(2)}(r)\,,
\eea
where the subscripts ``${\rm SD}$'' and ``${\rm SI}$'' stand for ``spin-dependent'' and ``spin-independent'' respectively. 
$V^{(0)}_{\kappa\lambda}(r)$ is the static potential
\bea
V^{(0)}_{\kappa\lambda}(r)=\Lambda_{\kappa}+V^{(0)}_o(r)+\dots\,,
\eea
where $\Lambda_{\kappa}$ is the gluelump energy, computable on the lattice~\cite{Juge:2002br}, and $V^{(0)}_o(r)$ is the static octet potential in the Lagrangian of weakly-coupled pNRQCD.
In this paper, we will only consider the lowest-lying gluelumps $\kappa=1^{+-}$, for which we will simply write subscripts $\kappa$ as $1$. The spin-dependent potentials have the form
\bea
V_{1\la\lap\,{\rm SD}}^{(1)}(r)&=&V_{1\,{\rm SK}}(r)\left(P^{i\dag}_{1\la}\bm{K}^{ij}_{1}P^j_{1\lap}\right)\cdot\bm{S}\nonumber\\
&&\, + V_{{ 1\,{\rm SK}}b}(r)\left[\left(\bm{r}\cdot P^{\dag}_{1\la}\right)\left(r^i\bm{K}^{ij}P^j_{1\lap}\right)\cdot\bm{S}
+\left(r^i\bm{K}^{ij}P^{j\dag }_{1\la}\right)\cdot\bm{S} \left(\bm{r}\cdot P_{1\lap}\right)\right]+\dots
\,,\label{sdm}\\
V_{1\la\lap\,{\rm SD}}^{(2)}(r)&=&V_{1\,{\rm SL}a}(r)\left(P^{i\dag}_{1\la}\bm{L}_{Q\bar{Q}}P^i_{1\lap}\right)\cdot\bm{S}
+V_{1\,{\rm SL}b}(r)P^{i\dag}_{1\la}\left(L_{Q\bar{Q}}^iS^j+S^iL_{Q\bar{Q}}^j\right)P^{j}_{1\lap}\nonumber\\
&&\, +V_{1\,{\rm S}^2}(r)\bm{S}^2\de_{\la\lap}+V_{1\,{\rm S}_{12}a}(r)S_{12}\de_{\la\lap}+V_{1\,{\rm S}_{12}b}(r)P^{i\dag}_{1\la}P^j_{1\lap}\left(S^i_1S^j_2+S^i_2S^j_1\right)+\dots\,,\,\,\,\,\,\,\,\,\,\,\label{sdm2}
\eea
where $\bm{L}_{Q\bar{Q}}$ is the orbital angular momentum of the heavy-quark-antiquark pair, 
$\bm{S}_1$ and $\bm{S}_2$ are the spin vectors of the heavy quark and heavy antiquark respectively,
$\bm{S}=\bm{S}_1+\bm{S}_2$ and ${S}_{12}=12(\bm{S}_1\cdot\hat{\bm{r}})(\bm{S}_2\cdot\hat{\bm{r}})-4\bm{S}_1\cdot\bm{S}_2$. 
$\left({K}^{ij}\right)^k=i\epsilon^{ijk}$ is the angular momentum operator for the spin-1 gluonic degrees of freedom. The projectors $P^{i}_{1\lambda}$ are given by
$P^{i}_{10}=\hat{r}_0^i= \hat{r}^i$, $
P^{i}_{1\pm 1}=\hat{r}^i_{\pm}=\mp\left(\hat{\theta}^i\pm i\hat{\phi}^i\right)/\sqrt{2}$.
The ellipses in Eqs.~(\ref{sdm}) and~(\ref{sdm2}) stand for terms suppressed by powers of $r\Lambda_{\rm QCD}$. It should be noted 
from Eq.~(\ref{sdm}) that the spin-dependent potential for quarkonium hybrids appears at order $1/m$, as opposed to traditional quarkonia, for
which the spin-dependent potential enters at order $1/m^2$. In addition, owing to the cylindrical symmetry in quarkonium
hybrids, the $1/m^2$-spin-dependent potential in Eq.~(\ref{sdm2}) has much involved structure than that of traditional quarkonia, which
has rotational symmetry instead.
In the matching, the object to consider is the two-point function:
\bea
&&I_{\kappa\la\la'}(\bm{r},\bm{R},\bm{r}',\bm{R}')\nn\\
&=&\lim_{T\to\infty}
\langle 0|P_{\kappa\la}^{i\dag} G^{ia\dag}_\kappa(\bm{R},T/2)O^a(\bm{r},\bm{R},T/2)O^{b\dag}(\bm{r}',\bm{R}',-T/2)P_{\kappa\la'}^j G^{jb}_\kappa(\bm{R}',-T/2)|0\rangle\,.
\label{eq:I_pNRQCD_1}
\eea
The matching is schematically depicted in Fig.~\ref{match}. In the figure, the single, double and curly lines represent the heavy-quark singlet, heavy-quark octet and gluon fields respectively. The black dots stand for vertices from weakly-coupled pNRQCD that involve a chromoelectric or chromomagnetic field.
The shaded blobs represent the nonperturbative gluon dynamics. 
Diagram (a) implies that the sum of the perturbative octet potential in pNRQCD and the gluelump energy appears in the quarkonium hybrid EFT. Diagrams (b)-(g) involve
insertions of gluon field strengths, which give rise to nonperturbative gluonic correlators. Denote the nonperturbative part of the coefficients
$V_i(r)$ on the right-hand side of Eqs.~(\ref{sdm}) and~(\ref{sdm2}) by
$V_i^{\rm np}(r)$. Then from the multipole expanion we know that $V_i^{\rm np}(r)$ can be expanded as $V_i^{\rm np}(r)=V_i^{{\rm np}\,(0)}
+V_i^{{\rm np}\,(1)}r^2+\dots$. Therefore, working to leading order and next-to-leading order in the multipole expansion for the $1/m^2$- and $1/m$-
terms respectively, we arrive at $
V_{\rm SK}(r)=V^{{\rm np}\,(0)}_{\rm SK}+V^{{\rm np}\,(1)}_{\rm SK}r^2,
V_{{\rm SK}b}(r)=V^{{\rm np}\,(0)}_{{\rm SK}b},
V_{{\rm SL}a(r)}=V_{o\,{\rm SL}}(r)+V^{{\rm np}\,(0)}_{{\rm SL}a},
V_{{\rm SL}b}(r)=V^{{\rm np}\,(0)}_{{\rm SL}b},
V_{{\rm S}^2}(r)=V_{o\,{\rm S}^2}(r)+V^{{\rm np}\,(0)}_{{\rm S}^2},
V_{{\rm S}_{12}a}(r)=V_{o\,{\rm S}_{12}}(r), 
V_{{\rm S}_{12}b}(r)=V^{{\rm np}\,(0)}_{{\rm S}_{12}b},
$
where $V_{o\,{\rm SL}}(r)$, $V_{o\,{\rm S}^2}(r)$ and $V_{o\,{\rm S}_{12}}(r)$ are the perturbative spin-dependent octet potential in pNRQCD. $V_i^{{\rm np}\,(j)}$ can be
expressed as a product of a perturbative coefficient and a gluonic correlator. For example,
diagram (b) in Fig.~\ref{match} involves an insertion of the 
$c_F\mathbf{S}\cdot\mathbf{B}/m$ vertex. This diagram gives 
\bea
V^{{\rm np}\,(0)}_{\rm SK}=c_F\tilde{U}^{\rm K}_{\rm B}\,,\label{eq:V_SK_factor}
\eea
where $c_F$ is a perturbative coefficient, and 
$\tilde{U}^{\rm K}_{\rm B}$ is a gluonic correlator:
\bea
\tilde{U}^K_B&=&\lim_{T\to\infty}\frac{ie^{i\Lambda_1}}{T}\frac{1}{12}\int^{T/2}_{-T/2}dt\langle 0|\bm{G}^{a \dag}(T/2)\cdot\left(g \bm{B}^{ac}_{{\rm adj}}(t)\times \bm{G}^c(-T/2)\right)|0\rangle\,,
\eea
with $\bm{B}_{{\rm adj}}^{ac}=if^{abc}\bm{B}^b$. The gluonic correlator can in principle be calculated on the lattice. Expressions of the other $V_i^{{\rm np}\,(j)}$ in terms of gluonic correlators
will be presented in~\cite{Long_spin}.

\section{Numerical results}\label{Results}
We obtain the spin splitting of the quarkonium hybrid spectrum by applying time-independent perturbation theory to the spin-dependent potentials in Eqs.~(\ref{sdm})-(\ref{sdm2}). We carry out perturbation theory to second order for the $V^{{\rm np}\,(0)}_{\rm SK}$-term in Eq.~(\ref{sdm}), and to first order for the $V^{{\rm np}\,(1)}_{\rm SK}$-term and $V^{{\rm np}\,(0)}_{{\rm SK}b}$-term in Eq.~(\ref{sdm}) and the $1/m^2$-suppressed operators in Eq.~(\ref{sdm2}). Note that in first-order perturbation theory the $V^{{\rm np}\,(0)}_{{\rm SK}b}$-term gives zero contribution.
The zeroth-order wavefunctions are obtained following the procedure described in Ref.~\cite{Berwein:2015vca}, by solving a set of coupled Schr\"odinger equations.
We will present the results for the four lowest-lying spin-multiplets shown in Table~\ref{tb:spin_multiplet}. In Table~\ref{tb:spin_multiplet}, $l(l+1)$ is the eigenvalue of $(\mathbf{L}_{Q\bar{Q}}+\mathbf{K})^2$.
\begin{table}[!t]
\caption{Lowest-lying quarkonium hybrid multiplets}
\begin{center}
\begin{tabular}{c|c|c|c}
\hline
\hline
Multiplet & $\,\,\,l\,\,\,$ & $J^{PC}(s=0)$ & $J^{PC}(s=1)$\\
\hline
$H_1$& $1$ & $1^{--}$ & $(0,1,2)^{-+}$ \\
$H_2$& $1$ & $1^{++}$ & $(0,1,2)^{+-}$ \\
$H_3$& $0$ & $0^{++}$ & $1^{+-}$ \\
$H_4$& $2$ & $2^{++}$ & $(1,2,3)^{+-}$ \\
\hline
\hline
\end{tabular}
\label{tb:spin_multiplet}
\end{center}
\end{table}
The six nonperturbative parameters $V^{{\rm np}\,(0)}_{\rm SK}$, $V^{{\rm np}\,(1)}_{\rm SK}$, $V^{{\rm np}\,(0)}_{{\rm SL}a}$, $V^{{\rm np}\,(0)}_{{\rm SL}b}$, $V^{{\rm np}\,(0)}_{{\rm S}^2}$, $V^{{\rm np}\,(0)}_{{\rm S}_{12}b}$ that appear in the spin-dependent potentials are obtained by fitting the charmonium hybrid spectrum obtained from our calculation to the lattice data from Ref.~\cite{Cheung:2016bym}, in which a a pion mass of $m_{\pi}\approx 240$ MeV is used. We take the values $m^{RS}_c(1{\rm GeV})=1.477$~GeV \cite{Pineda:2001zq} and $\alpha_s$ at $4$-loops with three light flavors, $\alpha_s(2.6\textrm{~GeV})=0.26$. In the fit, the lattice data are weighed by $(\Delta^2_{\textrm{lattice}}+\Delta^2_{\textrm{high-order}})^{-1/2}$, where $\Delta_{\textrm{lattice}}$ is the uncertainty of the lattice data and $\Delta_{\textrm{high-order}}=(m_{\textrm{lattice}}
-m_{\textrm{lattice spin-average}})\times\Lambda_{\rm QCD}/m$ is the estimated error due to higher-order terms in the potential. The $V^{\rm np}$'s in units of their natural size as powers of $\Lambda_{\rm QCD}$ are introduced to the fit through a prior. We take $\Lambda_{\rm QCD}=0.5$~GeV. 
\begin{figure*}[!t]
\begin{center}
\includegraphics[height=0.20\textheight,width=0.35\textwidth]{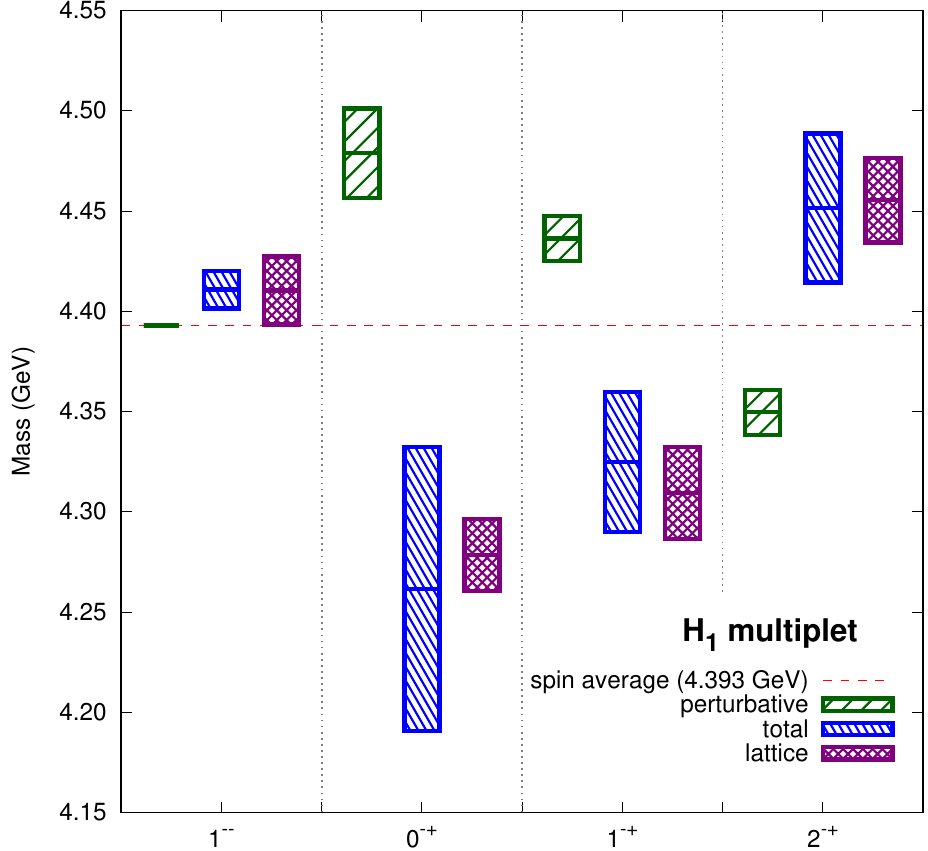}
\hspace*{0.50cm}
\includegraphics[height=0.20\textheight,width=0.35\textwidth]{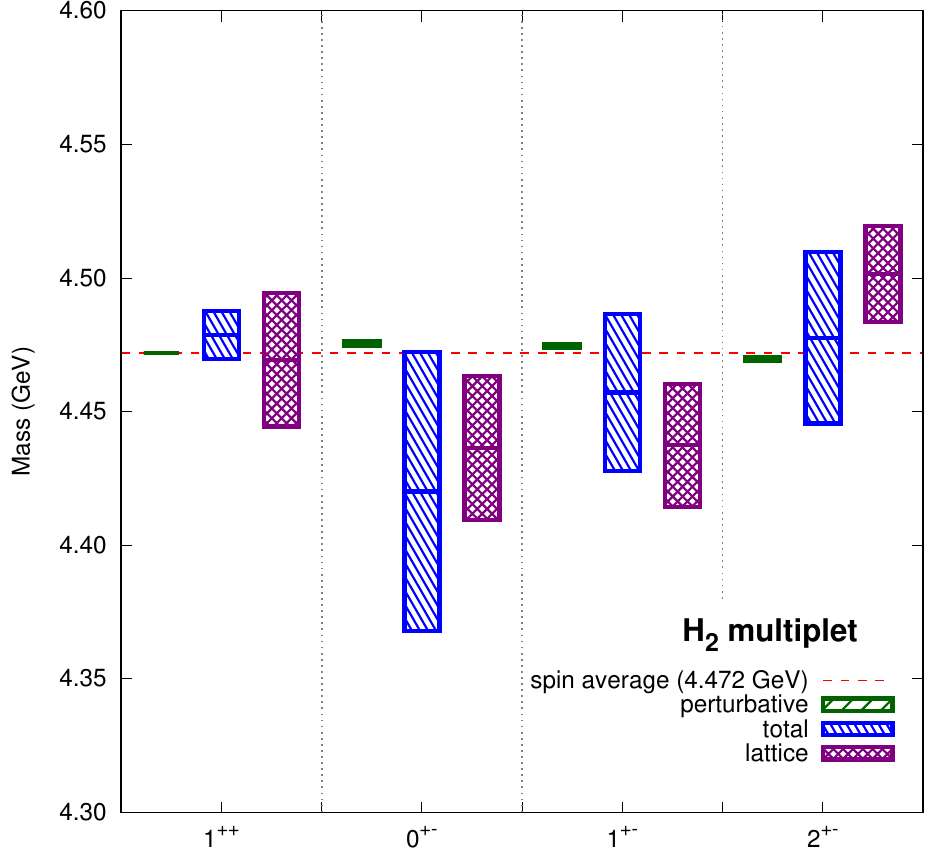} 
\\[2ex]
\includegraphics[height=0.20\textheight,width=0.35\textwidth]{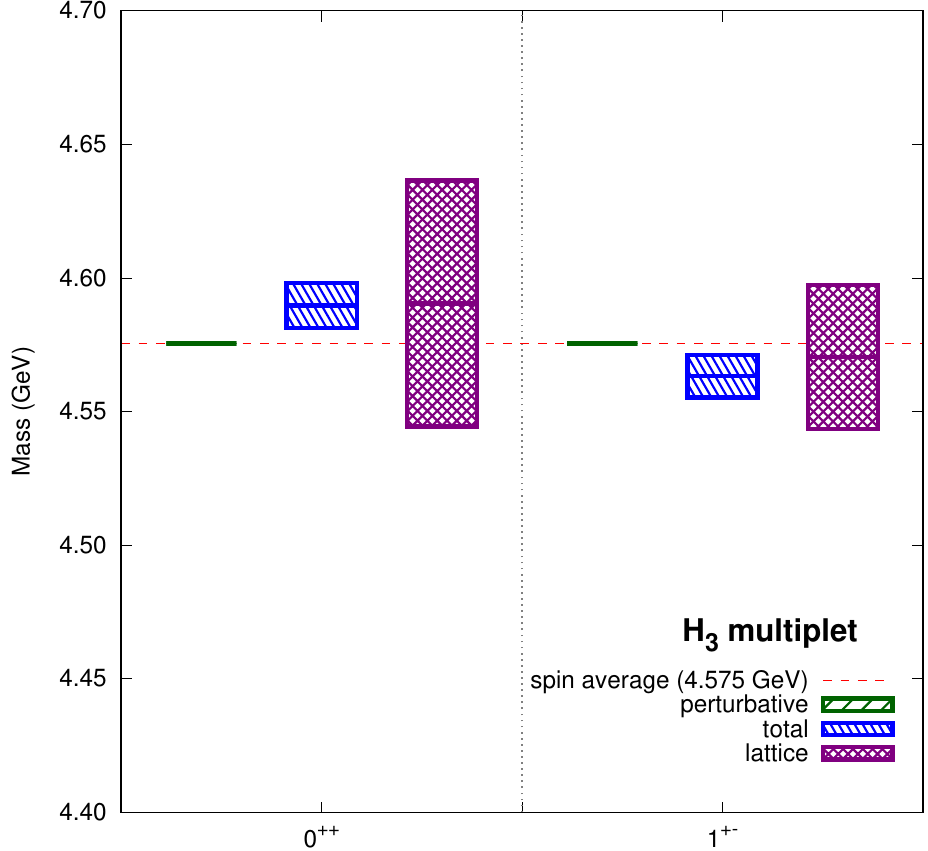}
\hspace*{0.50cm}
\includegraphics[height=0.20\textheight,width=0.35\textwidth]{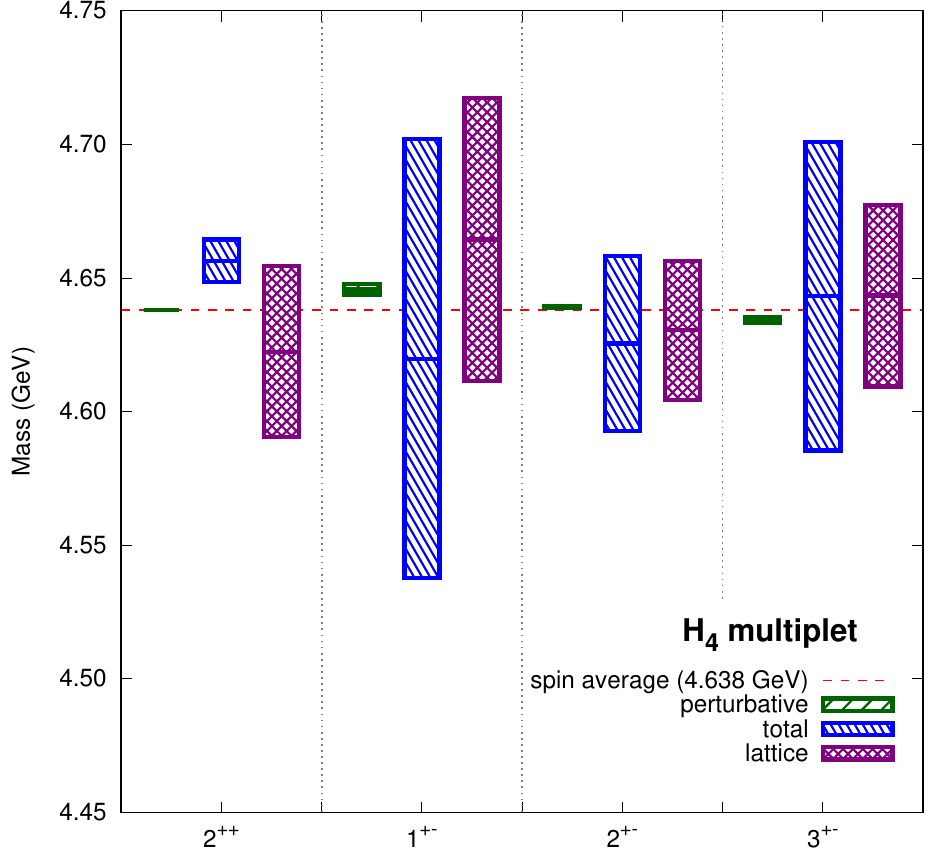}
\caption{Spectrum of the four lowest-lying charmonium hybrid multiplets. The lattice results from Ref.~\cite{Cheung:2016bym} are plotted in purple. In green we plotted the perturbative contributions to the spin-dependent operators in Eq.~(\ref{sdm2}) added to the spin average of the lattice results (red dashed line). In blue we show the full result of the spin-dependent operators of Eqs.~(\ref{sdm})-(\ref{sdm2}) including perturbative and nonperturbative contributions. The unknown nonperturbative matching coefficients are fitted to reproduce the lattice data. The height of the boxes indicate the uncertainty as detailed in the text.}
\label{fg:ccg_Cheung}
\end{center}
\end{figure*}
\begin{table}[!t]
\caption{Nonperturbative matching coefficients determined by fitting the charmonium hybrid spectrum obtained from the hybrid EFT to the lattice data from~\cite{Cheung:2016bym}. The matching coefficients are normalized to their parametrically natural sizes. We take the value $\Lambda_{QCD}=0.5$ GeV.}
\begin{center}
\begin{tabular}{c|c}\hline\hline
$V^{np\,(0)}_{SK}/\Lambda^2_{QCD}$       & $+1.03$\\
$V^{ np\,(1)}_{SK}/\Lambda^4_{QCD}$       & $-0.51$\\
$V^{ np\,(0)}_{SLa}/\Lambda^3_{QCD}$       & $-1.32$\\
$V^{np\,(0)}_{SLb}/\Lambda^3_{QCD}$       & $+2.44$\\
$V^{np\, (0)}_{S^2}/\Lambda^3_{QCD}$       & $-0.33$\\
$V^{np\,(0)}_{S_{12}b}/\Lambda^3_{QCD}$   & $-0.39$\\
\hline\hline
\end{tabular}
\label{tb:npfit}
\end{center}
\end{table}
The results of the fit are shown in Fig.~\ref{fg:ccg_Cheung}. The obtained values of the nonperturbative parameters $V^{\rm np}$'s are shown in Table~\ref{tb:npfit}. Each panel in Fig.~\ref{fg:ccg_Cheung} corresponds to one of the multiplets of Table~\ref{tb:spin_multiplet}. The purple boxes indicate the lattice results: the middle line corresponds to the mass of the state obtained from the lattice and the height of the box corresponds to the uncertainty. The red dashed line indicates the spin average mass of the lattice results. The green boxes correspond to the contribution to the spin splittings from the perturbative contributions. The height of the green box ($\Delta_{\rm p}$) is an estimate on the uncertainty given by the parametric size of higher order corrections, $\mathcal{O}(m\alpha_s^5)$. The blue boxes are the full results including the nonperturbative contributions after fitting the six nonperturbative parameters to the lattice data. The height of the blue box corresponds to the uncertainty of the full result. This uncertainty is given by $\Delta_{\rm full}=(\Delta_{\rm p}^2 +\Delta^2_{\rm np}+\Delta^2_{\rm fit})^{1/2}$, where the uncertainty of the nonperturbative contribution $\Delta_{\rm np}$ is estimated to be of parametric size of higher order corrections, $\mathcal{O}(\Lambda_{\rm QCD}(\Lambda_{\rm QCD}/m)^3)$, to the matching coefficients. $\Delta_{\rm fit}$ is the statistical error of the fit. 

An interesting feature is that for the spin triplets, the value of the perturbative contributions decreases with $J$. This trend is opposite to that of the lattice results. This discrepancy can be reconciled thanks to the nonperturbative contributions, in particular due to the contribution from $V^{{\rm np}\,(0)}_{\rm SK}$, which is of order $\Lambda_{\rm QCD}^2/m$ and is thus parametrically larger than the perturbative contributions, which are of order $mv^4$.
A consequence of the countervail of the perturbative contribution is a relatively large uncertainty on the full result caused by a large nonperturbative contribution. Due to this uncertainty the mass hierarchy among the spin triplet states of the multiplet $H_4$ is not firmly determined. However, as shown in Table~\ref{tb:npfit}, the values of the fitted parameters are consistent with the power counting of the EFT.

\begin{figure*}[!t]
\begin{center}
\includegraphics[height=0.20\textheight,width=0.35\textwidth]{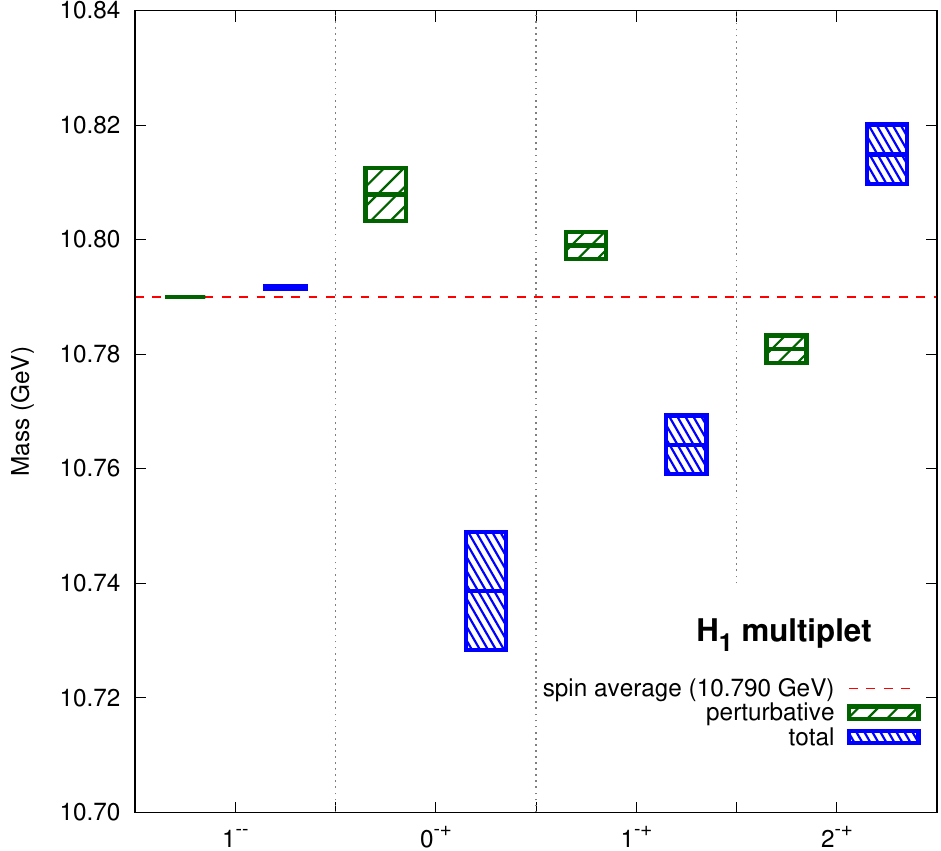}
\hspace*{0.50cm}
\includegraphics[height=0.20\textheight,width=0.35\textwidth]{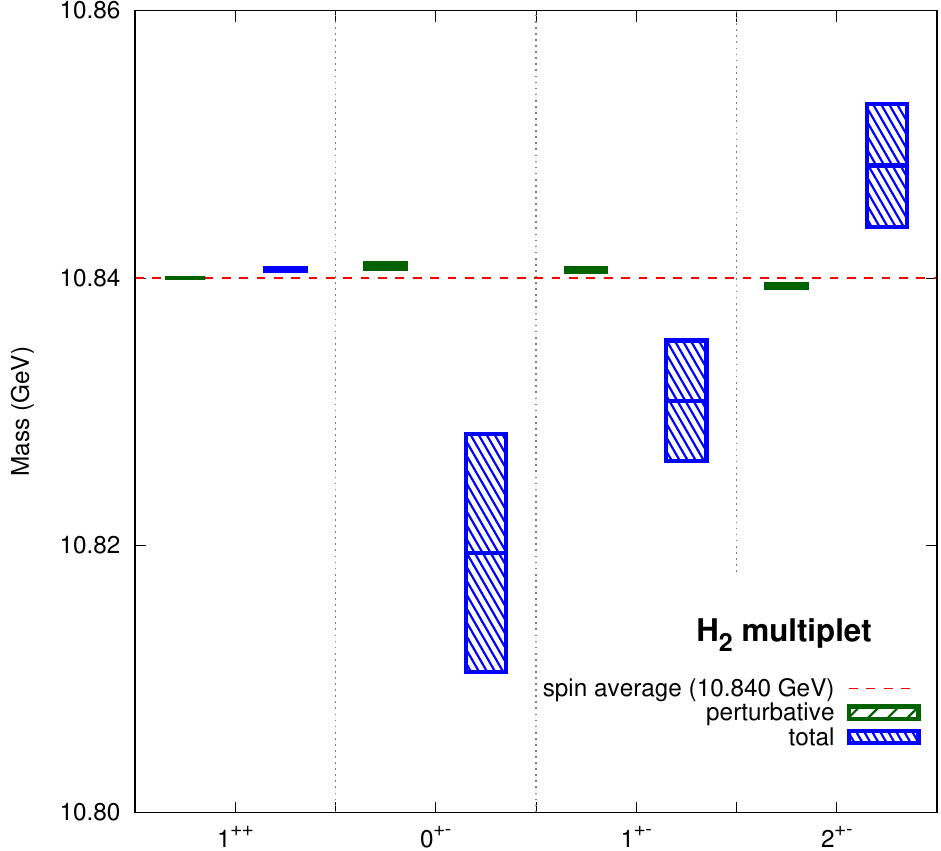} 
\\[2ex]
\includegraphics[height=0.20\textheight,width=0.35\textwidth]{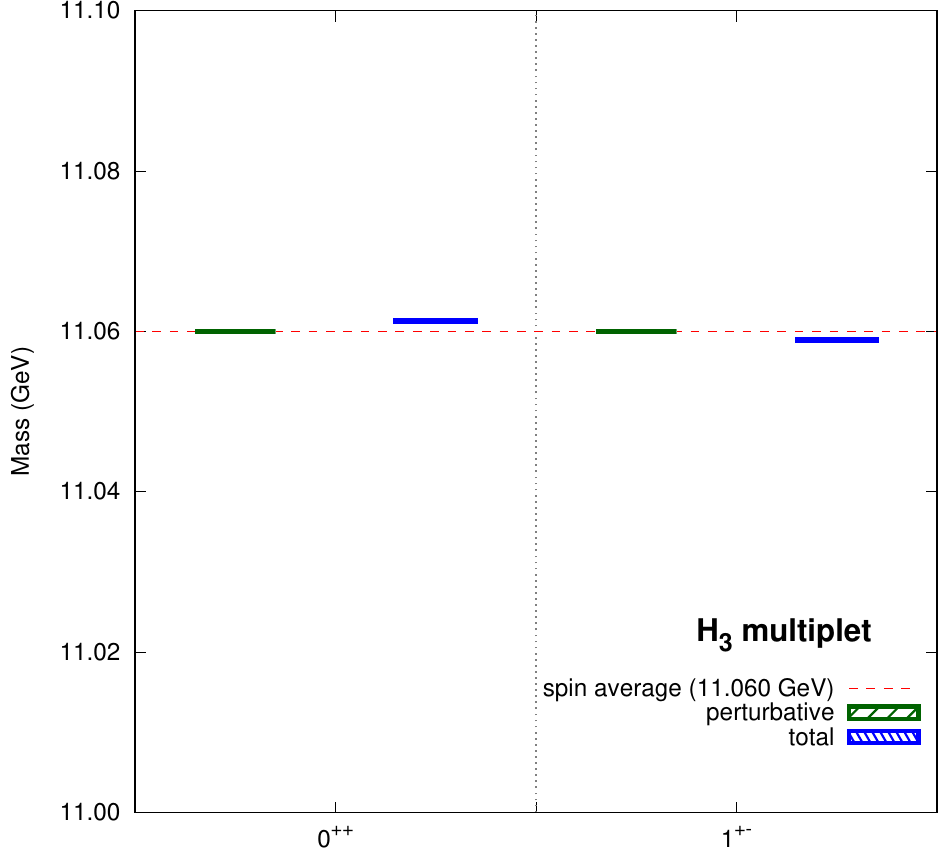}
\hspace*{0.50cm}
\includegraphics[height=0.20\textheight,width=0.35\textwidth]{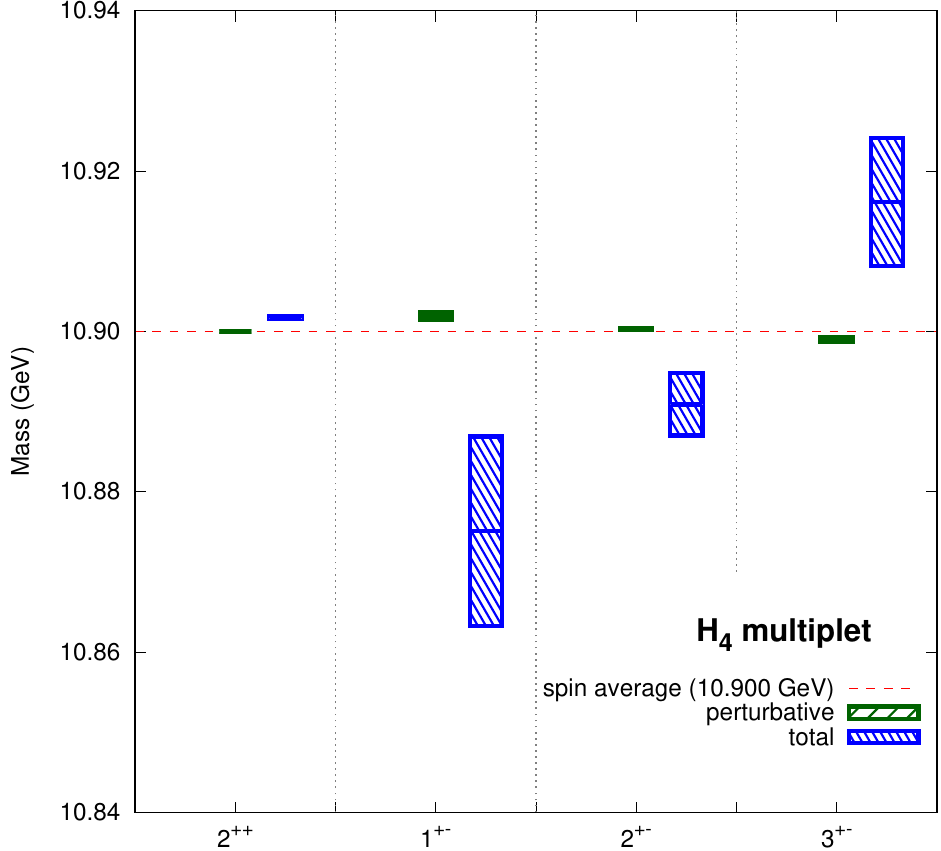}
\caption{Spectrum of the four lowest-lying bottomonium hybrids computed by adding the spin-dependent contributions to the spectrum obtained in Ref.~\cite{Berwein:2015vca}. The values of nonperturbative contribution to the matching coefficients are determined from the fit of the charmonium hybrids spectrum obtained from the hybrid EFT to the lattice data of Ref.~\cite{Cheung:2016bym} shown in Fig.~\ref{fg:ccg_Cheung}. The average mass for each multiplet is shown as a red dashed line. The results with only the perturbative contributions and the full results for the matching coefficients are shown as green and blue boxes respectively. The height of the boxes indicates the uncertainty as detailed in the text.}
\label{fig:bbg_Cheung}
\end{center}
\end{figure*}

Since the nonperturbative parameters $V^{\rm np}$'s are products of a perturbative coefficient, the flavor dependence of which is known, and a gluonic correlator, which is flavor-independent, we can 
use the obtained values of the $V^{\rm np}$'s from our fit to predict the bottomonium hybrid spectrum. We show the results thus obtained in Fig.~\ref{fig:bbg_Cheung}. At the order of accuracy we are working 
here, only $V^{{\rm np}\,(0)}_{\rm SK}$ among the six $V^{\rm np}$'s has flavor dependence, which is given by $c_F$ (Eq.~(\ref{eq:V_SK_factor})). The one-loop expression of $c_F$~\cite{Eichten:1990vp} is used. For the bottom mass here we use $m_b^{RS}(\textrm{1 GeV})=4.863$ GeV. 

\section{Conclusions}\label{Conclusions}
Under the framework of a nonrelativistic effective field theory, we derived for the first time the spin-dependent part of 
the heavy-quark-antiquark potential for heavy quarkonium hybrids to order $1/m^2$ in the heavy-quark-mass expansion. We found that 
several operators that are not found in standard quarkonia appear. Most notable is the operator that couples the spin of the heavy-quark-antiquark pair to the 
angular momentum of the gluonic excitation. This operator is suppressed only by $1/m$, as opposed to the case of traditional quarkonia, for which the spin-dependent
potential enters at order $1/m^2$. For the spin-dependent potential at $1/m^2$, the structure is more involved than the case of traditional quarkonia, owing to 
the fact that quarkonium hybrids possess cylindrical symmetry, while traditional quarkonia possess rotational symmetry. Under the approximation of short interquark distance $r\ll 1/\Lambda_{\rm QCD}$, we performed the matching between weakly-coupled pNRQCD and the quarkonium hybrid EFT. We showed that
in the EFT, the nonperturbative part of the matching coefficients in the spin-dependent potential are factorized into a perturbative factor and a nonperturbative gluonic correlator.

Using time-independent perturbation theory, we computed the quarkonium hybrid spectrum using the quarkonium hybrid EFT, treating the spin-independent potential as a perturbation. Six nonperturbative
parameters that appear in the spin-dependent potential were fitted to the lattice data of the charmonium hybrid spectrum. We found that the perturbative
contributions have a trend opposite to the lattice data, and the inclusion of nonperturbative contributions is necessary. Moreover, the nonperturbative contributions, most of which come from the $1/m$-suppressed operator, dominate over the perturbative contributions. We also found that the fitted values of the nonperturbatibe parameters are consistent with the power counting of the EFT.
With the fitted values of the nonperturbative parameters, we predicted the bottomonium hybrid spectrum, for which lattice calculations are still sparse.

As a final remark, we note that, since this framework of EFT can be generalized to describe states with light degrees of freedom other than gluons, 
such as heavy tetraquarks and pentaquarks~\cite{Braaten:2014ita,Brambilla:2017uyf}, 
a similar analysis could be done to describe spin multiplets of exotic quarkonia other than hybrids,
eventually providing an unified description of all heavy-quark-antiquark spin multiplets.
\section*{Acknowledgements}
We thank Nora Brambilla, Jorge Segovia, Jaume Tarr\'us Castell\`a, and Antonio Vairo for advice and collaboration on this work.
This work has been supported by the DFG and the NSFC through funds 
provided to the Sino-German CRC 110 ``Symmetries and the Emergence of Structure in QCD'', 
and by the DFG cluster of excellence ``Origin and structure of the universe'' (www.universe-cluster.de). 

\bibliographystyle{JHEP}
\bibliography{hybridspin}

\providecommand{\href}[2]{#2}\begingroup\raggedright\begin{thebibliography}{10}

\bibitem{Choi:2003ue}
{\scshape Belle} collaboration, S.~K. Choi et~al., \emph{{Observation of a
  narrow charmonium - like state in exclusive B+- ---> K+- pi+ pi- J / psi
  decays}}, \href{https://doi.org/10.1103/PhysRevLett.91.262001}{\emph{Phys.
  Rev. Lett.} {\bfseries 91} (2003) 262001}
  [\href{https://arxiv.org/abs/hep-ex/0309032}{{\ttfamily hep-ex/0309032}}].

\bibitem{Brambilla:2004wf}
{\scshape Quarkonium Working Group} collaboration, N.~Brambilla et~al.,
  \emph{{Heavy quarkonium physics}},
  \href{https://arxiv.org/abs/hep-ph/0412158}{{\ttfamily hep-ph/0412158}}.

\bibitem{Brambilla:2010cs}
N.~Brambilla et~al., \emph{{Heavy quarkonium: progress, puzzles, and
  opportunities}},
  \href{https://doi.org/10.1140/epjc/s10052-010-1534-9}{\emph{Eur. Phys. J.}
  {\bfseries C71} (2011) 1534}
  [\href{https://arxiv.org/abs/1010.5827}{{\ttfamily 1010.5827}}].

\bibitem{Brambilla:2014jmp}
N.~Brambilla et~al., \emph{{QCD and Strongly Coupled Gauge Theories: Challenges
  and Perspectives}},
  \href{https://doi.org/10.1140/epjc/s10052-014-2981-5}{\emph{Eur. Phys. J.}
  {\bfseries C74} (2014) 2981}
  [\href{https://arxiv.org/abs/1404.3723}{{\ttfamily 1404.3723}}].

\bibitem{Olsen:2014qna}
S.~L. Olsen, \emph{{A New Hadron Spectroscopy}},
  \href{https://doi.org/10.1007/S11467-014-0449-6}{\emph{Front. Phys.(Beijing)}
  {\bfseries 10} (2015) 121} [\href{https://arxiv.org/abs/1411.7738}{{\ttfamily
  1411.7738}}].

\bibitem{Dudek:2007wv}
J.~J. Dudek, R.~G. Edwards, N.~Mathur and D.~G. Richards, \emph{{Charmonium
  excited state spectrum in lattice QCD}},
  \href{https://doi.org/10.1103/PhysRevD.77.034501}{\emph{Phys. Rev.}
  {\bfseries D77} (2008) 034501}
  [\href{https://arxiv.org/abs/0707.4162}{{\ttfamily 0707.4162}}].

\bibitem{Bali:2011dc}
G.~Bali et~al., \emph{{Spectra of heavy-light and heavy-heavy mesons containing
  charm quarks, including higher spin states for $N_f=2+ 1$}},
  \href{https://doi.org/10.22323/1.139.0135}{\emph{PoS} {\bfseries LATTICE2011}
  (2011) 135} [\href{https://arxiv.org/abs/1108.6147}{{\ttfamily 1108.6147}}].

\bibitem{Bali:2011rd}
G.~S. Bali, S.~Collins and C.~Ehmann, \emph{{Charmonium spectroscopy and mixing
  with light quark and open charm states from $n_F$=2 lattice QCD}},
  \href{https://doi.org/10.1103/PhysRevD.84.094506}{\emph{Phys. Rev.}
  {\bfseries D84} (2011) 094506}
  [\href{https://arxiv.org/abs/1110.2381}{{\ttfamily 1110.2381}}].

\bibitem{Liu:2012ze}
{\scshape Hadron Spectrum} collaboration, L.~Liu, G.~Moir, M.~Peardon, S.~M.
  Ryan, C.~E. Thomas, P.~Vilaseca et~al., \emph{{Excited and exotic charmonium
  spectroscopy from lattice QCD}},
  \href{https://doi.org/10.1007/JHEP07(2012)126}{\emph{JHEP} {\bfseries 07}
  (2012) 126} [\href{https://arxiv.org/abs/1204.5425}{{\ttfamily 1204.5425}}].

\bibitem{Cheung:2016bym}
{\scshape Hadron Spectrum} collaboration, G.~K.~C. Cheung, C.~O'Hara, G.~Moir,
  M.~Peardon, S.~M. Ryan, C.~E. Thomas et~al., \emph{{Excited and exotic
  charmonium, $D_s$ and $D$ meson spectra for two light quark masses from
  lattice QCD}}, \href{https://doi.org/10.1007/JHEP12(2016)089}{\emph{JHEP}
  {\bfseries 12} (2016) 089}
  [\href{https://arxiv.org/abs/1610.01073}{{\ttfamily 1610.01073}}].

\bibitem{Juge:1997nc}
K.~J. Juge, J.~Kuti and C.~J. Morningstar, \emph{{Gluon excitations of the
  static quark potential and the hybrid quarkonium spectrum}},
  \href{https://doi.org/10.1016/S0920-5632(97)00759-7}{\emph{Nucl. Phys. Proc.
  Suppl.} {\bfseries 63} (1998) 326}
  [\href{https://arxiv.org/abs/hep-lat/9709131}{{\ttfamily hep-lat/9709131}}].

\bibitem{Bali:2000vr}
{\scshape TXL, T(X)L} collaboration, G.~S. Bali, B.~Bolder, N.~Eicker,
  T.~Lippert, B.~Orth, P.~Ueberholz et~al., \emph{{Static potentials and
  glueball masses from QCD simulations with Wilson sea quarks}},
  \href{https://doi.org/10.1103/PhysRevD.62.054503}{\emph{Phys. Rev.}
  {\bfseries D62} (2000) 054503}
  [\href{https://arxiv.org/abs/hep-lat/0003012}{{\ttfamily hep-lat/0003012}}].

\bibitem{Juge:2002br}
K.~J. Juge, J.~Kuti and C.~Morningstar, \emph{{Fine structure of the QCD string
  spectrum}}, \href{https://doi.org/10.1103/PhysRevLett.90.161601}{\emph{Phys.
  Rev. Lett.} {\bfseries 90} (2003) 161601}
  [\href{https://arxiv.org/abs/hep-lat/0207004}{{\ttfamily hep-lat/0207004}}].

\bibitem{Bali:2003jq}
G.~S. Bali and A.~Pineda, \emph{{QCD phenomenology of static sources and
  gluonic excitations at short distances}},
  \href{https://doi.org/10.1103/PhysRevD.69.094001}{\emph{Phys. Rev.}
  {\bfseries D69} (2004) 094001}
  [\href{https://arxiv.org/abs/hep-ph/0310130}{{\ttfamily hep-ph/0310130}}].

\bibitem{Griffiths:1983ah}
L.~A. Griffiths, C.~Michael and P.~E.~L. Rakow, \emph{{Mesons With Excited
  Glue}}, \href{https://doi.org/10.1016/0370-2693(83)90680-9}{\emph{Phys.
  Lett.} {\bfseries 129B} (1983) 351}.

\bibitem{Braaten:2014qka}
E.~Braaten, C.~Langmack and D.~H. Smith, \emph{{Born-Oppenheimer Approximation
  for the XYZ Mesons}},
  \href{https://doi.org/10.1103/PhysRevD.90.014044}{\emph{Phys. Rev.}
  {\bfseries D90} (2014) 014044}
  [\href{https://arxiv.org/abs/1402.0438}{{\ttfamily 1402.0438}}].

\bibitem{Braaten:2014ita}
E.~Braaten, C.~Langmack and D.~H. Smith, \emph{{Selection Rules for Hadronic
  Transitions of XYZ Mesons}},
  \href{https://doi.org/10.1103/PhysRevLett.112.222001}{\emph{Phys. Rev. Lett.}
  {\bfseries 112} (2014) 222001}
  [\href{https://arxiv.org/abs/1401.7351}{{\ttfamily 1401.7351}}].

\bibitem{Berwein:2015vca}
M.~Berwein, N.~Brambilla, J.~Tarr\'us~Castell\`a and A.~Vairo,
  \emph{{Quarkonium Hybrids with Nonrelativistic Effective Field Theories}},
  \href{https://doi.org/10.1103/PhysRevD.92.114019}{\emph{Phys. Rev.}
  {\bfseries D92} (2015) 114019}
  [\href{https://arxiv.org/abs/1510.04299}{{\ttfamily 1510.04299}}].

\bibitem{Oncala:2017hop}
R.~Oncala and J.~Soto, \emph{{Heavy Quarkonium Hybrids: Spectrum, Decay and
  Mixing}}, \href{https://doi.org/10.1103/PhysRevD.96.014004}{\emph{Phys. Rev.}
  {\bfseries D96} (2017) 014004}
  [\href{https://arxiv.org/abs/1702.03900}{{\ttfamily 1702.03900}}].

\bibitem{Brambilla:2017uyf}
N.~Brambilla, G.~o. Krein, J.~Tarr\'us~Castell\`a and A.~Vairo, \emph{{The
  Born-Oppenheimer approximation in an effective field theory language}},
  \href{https://arxiv.org/abs/1707.09647}{{\ttfamily 1707.09647}}.

\bibitem{Brambilla:2018pyn}
N.~Brambilla, W.~K. Lai, J.~Segovia, J.~Tarr\'us~Castell\`a and A.~Vairo,
  \emph{{Spin structure of heavy-quark hybrids}},
  \href{https://arxiv.org/abs/1805.07713}{{\ttfamily 1805.07713}}.

\bibitem{Long_spin}
N.~Brambilla, W.~K. Lai, J.~Segovia, J.~Tarr\'us~Castell\`a and A.~Vairo,
  {{TUM-EFT 96/17 (in preparation)}.} 

\bibitem{Caswell:1985ui}
W.~E. Caswell and G.~P. Lepage, \emph{{Effective Lagrangians for Bound State
  Problems in QED, QCD, and Other Field Theories}},
  \href{https://doi.org/10.1016/0370-2693(86)91297-9}{\emph{Phys. Lett.}
  {\bfseries 167B} (1986) 437}.

\bibitem{Bodwin:1994jh}
G.~T. Bodwin, E.~Braaten and G.~P. Lepage, \emph{{Rigorous QCD analysis of
  inclusive annihilation and production of heavy quarkonium}},
  \href{https://doi.org/10.1103/PhysRevD.55.5853,
  10.1103/PhysRevD.51.1125}{\emph{Phys. Rev.} {\bfseries D51} (1995) 1125}
  [\href{https://arxiv.org/abs/hep-ph/9407339}{{\ttfamily hep-ph/9407339}}].

\bibitem{Manohar:1997qy}
A.~V. Manohar, \emph{{The HQET / NRQCD Lagrangian to order alpha / m-3}},
  \href{https://doi.org/10.1103/PhysRevD.56.230}{\emph{Phys. Rev.} {\bfseries
  D56} (1997) 230} [\href{https://arxiv.org/abs/hep-ph/9701294}{{\ttfamily
  hep-ph/9701294}}].

\bibitem{Pineda:1997bj}
A.~Pineda and J.~Soto, \emph{{Effective field theory for ultrasoft momenta in
  NRQCD and NRQED}},
  \href{https://doi.org/10.1016/S0920-5632(97)01102-X}{\emph{Nucl. Phys. Proc.
  Suppl.} {\bfseries 64} (1998) 428}
  [\href{https://arxiv.org/abs/hep-ph/9707481}{{\ttfamily hep-ph/9707481}}].

\bibitem{Brambilla:1999xf}
N.~Brambilla, A.~Pineda, J.~Soto and A.~Vairo, \emph{{Potential NRQCD: An
  effective theory for heavy quarkonium}},
  \href{https://doi.org/10.1016/S0550-3213(99)00693-8}{\emph{Nucl. Phys.}
  {\bfseries B566} (2000) 275}
  [\href{https://arxiv.org/abs/hep-ph/9907240}{{\ttfamily hep-ph/9907240}}].

\bibitem{Foster:1998wu}
{\scshape UKQCD} collaboration, M.~Foster and C.~Michael, \emph{{Hadrons with a
  heavy color adjoint particle}},
  \href{https://doi.org/10.1103/PhysRevD.59.094509}{\emph{Phys. Rev.}
  {\bfseries D59} (1999) 094509}
  [\href{https://arxiv.org/abs/hep-lat/9811010}{{\ttfamily hep-lat/9811010}}].

\bibitem{Pineda:2001zq}
A.~Pineda, \emph{{Determination of the bottom quark mass from the Upsilon(1S)
  system}}, \href{https://doi.org/10.1088/1126-6708/2001/06/022}{\emph{JHEP}
  {\bfseries 06} (2001) 022}
  [\href{https://arxiv.org/abs/hep-ph/0105008}{{\ttfamily hep-ph/0105008}}].

\bibitem{Eichten:1990vp}
E.~Eichten and B.~R. Hill, \emph{{STATIC EFFECTIVE FIELD THEORY: 1/m
  CORRECTIONS}},
  \href{https://doi.org/10.1016/0370-2693(90)91408-4}{\emph{Phys. Lett.}
  {\bfseries B243} (1990) 427}.

\end{thebibliography}\endgroup

\end{document}